\documentclass[12pt]{iopart}
\usepackage{soul}
\usepackage{color}
\usepackage{todonotes}
\usepackage[colorlinks=true,citecolor=blue]{hyperref}
\usepackage[]{graphicx}
\usepackage{subcaption}
\usepackage{multirow}

\newcommand{\argmax}{\mathop{\mathrm{arg\,max}}}

\begin{document}

\title[]{Plasma Confinement Mode Classification Using a Sequence-to-Sequence Neural Network With Attention}

\author{F. Matos$^1$, V. Menkovski$^2$, A. Pau$^3$, G. Marceca$^3$, F. Jenko$^1$ and the TCV Team$^3$\footnote{See author list of S. Coda et al 2019 Nucl. Fusion 59 112023}}

\address{$^1$ Max Planck Institute for Plasma Physics, Boltzmannstr. 2, 85748 Garching, Germany}
\address{$^2$ Eindhoven University of Technology, 5612 AZ Eindhoven, Netherlands}
\address{$^3$ École Polytechnique Fédérale de Lausanne (EPFL), Swiss Plasma Center (SPC), CH-1015 Lausanne, Switzerland}

\ead{francisco.matos@ipp.mpg.de}

\vspace{10pt}

\begin{abstract}

In a typical fusion experiment, the plasma can have several possible confinement modes. At the TCV tokamak, aside from the Low (L) and High (H) confinement modes, an additional mode, dithering (D), is frequently observed. Developing methods that automatically detect these modes is considered to be important for future tokamak operation. Previous work\cite{matos2020classification} with deep learning methods, particularly convolutional recurrent neural networks (Conv-RNNs), indicates that they are a suitable approach. Nevertheless, those models are sensitive to noise in the temporal alignment of labels, and that model in particular is limited to making individual decisions taking into account only its own hidden state and its input at each time step. In this work, we propose an architecture for a sequence-to-sequence neural network model with attention which solves both of those issues. Using a carefully calibrated dataset, we compare the performance of a Conv-RNN with that of our proposed sequence-to-sequence model, and show two results: one, that the Conv-RNN can be improved upon with new data; two, that the sequence-to-sequence model can improve the results even further, achieving excellent scores on both train and test data. 

\end{abstract}

\vspace{2pc}
\noindent{\it Keywords}: CNN, LSTM, Deep Learning, H mode, L mode, Dither, Sequence-to-Sequence, Attention

\section{Introduction}

During nuclear fusion experiments, the plasma can be described as being in one of several possible confinement states. At the TCV tokamak, it is typically classified as being in either Low(L), Dithering(D) or High(H) confinement mode. All shots, during the ramp-up phase of the plasma, begin in L mode. By applying sufficient heating power, the plasma spontaneously transitions into H mode\cite{xu2014dynamics} (typically at TCV this process lasts approximately $1ms$). This mode is termed High confinement because, once it is reached, one can observe significantly reduced transport of particles and energy from the plasma to the surrounding vessel walls. This allows for a larger energy confinement per input power; for this reason, most current designs for future tokamaks assume that they will regularly run in H-mode. In some cases the transition from L to H mode does not happen directly, but rather the plasma oscillates rapidly between the two confinement regimes. In this case, the plasma is considered to be in a Dithering\cite{nielsen2015simulation} mode. 

Many studies have been done on the physical factors behind the transition between L and H mode, but the phenomenon is still not completely understood\cite{martin2008power}. Furthermore, there is no simple set of rules that can used to determine the plasma mode given the values of the signals of fusion experiment. Nevertheless, most of the time, there are highly salient patterns in these measured signals that can be used by domain experts to determine the plasma mode with high confidence. For example, a transition from L to H mode can typically be identified by observing a sudden drop in the emitted plasma radiation. However, these data patterns can be rather complicated and ambiguous; for example, Dithers leave signatures in the emission of photons similar to that of type III Edge Localized Modes\cite{ryter1994h}, which are events that occur during H-mode.

This process of manually labeling the experimental data can be quite cumbersome in many cases, particularly when one wishes to conduct large studies and analyze many shots. For that reason, work has been put into developing tools capable of automating the task of detecting different confinement modes. In particular, in the past few years, research has been done on using Machine Learning\cite{vega2009automated, gonzalez2012automatic, murari2006fuzzy, lukianitsa2008analyses, meakins2010application} and, more recently, Deep Learning\cite{matos2020classification} for this task. These algorithms are particularly suitable for dealing with such challenges of extracting patterns of such high-dimensional data collected during these experiments.

For example, when analyzing transitions between L and H modes, the type of correlations one expects to find in the data --- localized, spatial correlations as well as long-term temporal ones --- can be, respectively, efficiently discovered using Convolutional\cite{krizhevsky2012imagenet, ciregan2012multi} and Recurrent Neural Networks (RNNs)\cite{graves2013speech, elman1990finding}. Previous work with these models indicates that they can be very accurate in this task\cite{matos2020classification}. One of the main challenges of these models, however, is that they have to produce a decision about the plasma mode at each time step by looking only at a given context of the signals and their own past states. In contrast, when a human expert faces a difficult decision, they regularly reason through several possible sequences of mode evolutions. They go back and forward through the input signal, consider the consequences of labeling a mode with a given value for all consecutive modes, and in doing so, frequently revise decisions until the most likely sequence of plasma confinement modes can be determined. 
The convolutional RNN model, by itself, is incapable of producing decisions over sequences of outputs and is limited to making a sequence of individual decisions. Furthermore, a vanilla RNN is susceptible to noise produced by misaligned labels. A class of more powerful models, the sequence-to-sequence models, can solve both of these problems. These models, as well as associated mechanisms such as attention\cite{bahdanau2014neural, luong2015effective, xu2015show}, have considerably advanced the field of neural machine translation and transduction in the past few years.

Conceptually, the tasks of automated language translation and the automated labeling of plasma confinement modes are closely related: one wishes to translate a sentence in a source language to a different sentence with the same meaning in a target language. In the case of automated labeling of plasma confinement modes, one can consider signal time traces to constitute the sentence in the source language, while the corresponding confinement modes can be thought of as the ``translated'' sentence in the target language. For this reason, in this paper, we propose an approach that builds upon previous work with deep learning applied to automated detection of plasma confinement modes, by using recent developments in the field of neural machine translation.

Section \ref{sec:background} provides an overview of the field of neural machine translation, in particular by explaining the functioning of sequence-to-sequence models, and how we can adapt them to suit our task. Section \ref{sec:arch} details our considerations regarding the data and the problem formulation, our preprocessing steps and the proposed model architecture. Section \ref{sec:results} shows some of the obtained results and scores, and in particular, we compare the results of this model with those obtained in\cite{matos2020classification}. We then wrap up with a discussion in section \ref{sec:discussion}. 

\section{Background}\label{sec:background}

\subsection{Sequence-to-sequence models}\label{subsec:seqtoseq}

Sequence-to-sequence models have achieved tremendous success in the field of neural machine translation\cite{sutskever2014sequence}. These models are characterized by two separate components performing different tasks: an \textit{encoder} that reads a sentence in the source language and produces an encoded representation of that sentence, and a \textit{decoder} that, based on the \textit{encoding}, produces an appropriate translation into the target language. The encoder and decoder can technically be any type of algorithm, though in most applications, they are built with RNNs\cite{cho2014learning} that are jointly trained.

In neural machine translation, the encoder typically maps a word or sequence of words in the source language to a numerical representation of said words, as a function of a pre-defined size of the source language vocabulary. This is then followed by a recurrent layer, typically a long short-term memory (LSTM) layer, or a gated recurrent unit (GRU) that is trained to find sequential correlations in the embedded input sentences. These models can keep track of long-term temporal correlations because they have internal \textit{hidden states}. At every source timestep $j$, with $0 < j < k$ (where $k$ is the length of the source sequence), the encoder computes a new hidden state, $h_j$; each vector $h_j$ in the sequence of hidden states $(h_0, ..., h_k)$ constitutes a summary of the information that the encoder has processed until timestep $j$, and the final hidden state of the encoder's recurrent layer ($h_k$) can thus be considered to be an encoded representation of the entire input sequence.

The decoder must, subject to the encoding produced by the encoder ($h_k$), produce an appropriate corresponding sequence of words in the target language. When using an RNN decoder, this is done by setting its initial hidden state to $h_k$. Therefore, unlike a simple RNN model, which only receives a part of the source data as input at each time step, in the sequence-to-sequence model, the decoder works with a representation of the entire source sequence. The decoder then outputs, at each decoding timestep, a probability distribution over the discrete set of possible outputs, conditioned on the source sequence. Furthermore, in the general formulation of the sequence modeling problem, where the input sequence is not aligned to the output (e.g., different sampling rates of the input and output), the model needs not only to determine the output sequence, but also to align the symbols of the output sequence with the input. Lastly, the decoder can be made autoregressive, by feeding it a selected output at the previous timestep as an input in the next.

\begin{figure*}[h!]
    \centering
        \includegraphics[scale=0.5, trim=60 260 65 30, clip]{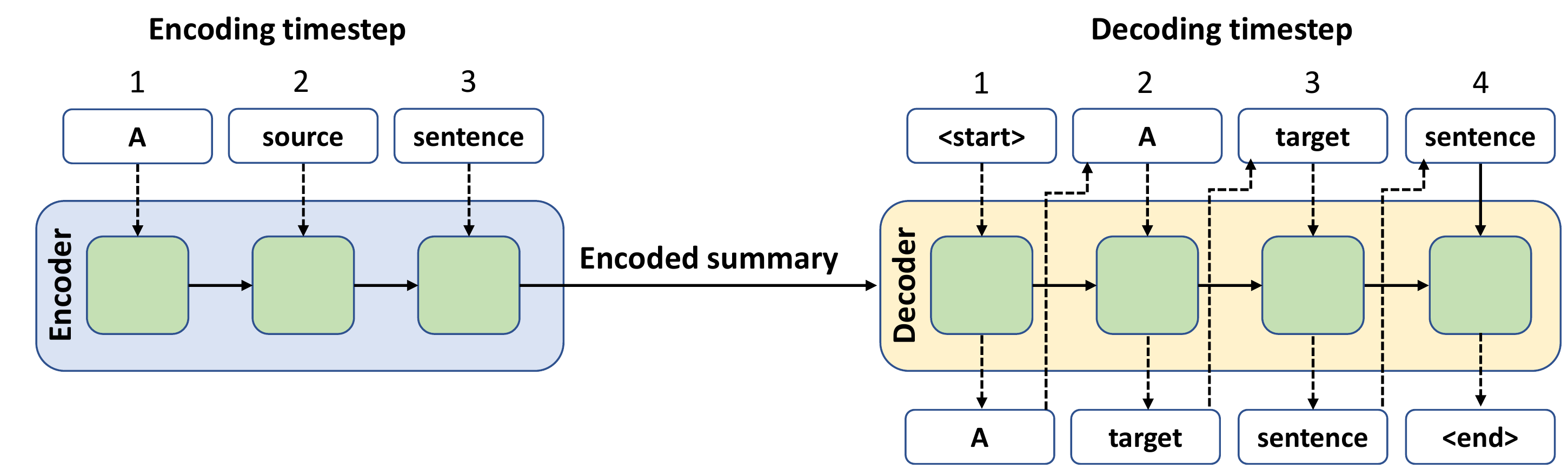}
        \caption[Flow of information in a sequence-to-sequence encoder-decoder model]{Representation of the flow of information in a sequence-to-sequence encoder-decoder model. Dashed lines denote model inputs and outputs, while solid lines denote hidden states. At decoding timestep $1$, the output distribution gives highest probability to the word $A$, which is then fed as input at timestep $2$. }
        \label{fig:encoder-decoder}
\end{figure*}

Figure \ref{fig:encoder-decoder} illustrates this mechanism: the encoder produces an encoding of the input sequence; the decoder, using that encoding as a starting state, produces a translation into a target sentence. At time step $1$ of the decoding process, a $<$start$>$ character is fed to the decoder; in subsequent steps, the selected output from the previous time step is fed as input.

An autoregressive decoder can, at inference time, evaluate several output sequences in parallel, and therefore is not limited to outputting a single solution, but rather, it can produce a probability distribution of possible solutions. In practice, sampling from the output joint probability distribution can be done by treating the distribution as a tree data structure, where each path in the tree represents a sample from the distribution, i.e., a different possible output sequence. Expanding a new node in that tree corresponds to sampling a different output from the previous timestep to condition future outputs; searching this tree for solutions (paths) of high probability can be done efficiently with a beam search algorithm\cite{graves2012sequence}. In addition, the tree search can be done (and simplified) by explicitly incorporating domain knowledge, which can allow for pruning paths that are known to be impossible (for example, in our case, by discarding paths that start in H-mode). All of this contrasts with using a simple RNN for translation. In that case, one expects the model to output a (single) solution of high likelihood, there is no way (in inference time) to explicitly encode knowledge that rules out certain solutions, and the source and target must be of the same length.

\subsection{Attention}
\label{subseq:attention}

While encoder-decoder models achieve very good results, they can nevertheless still lose performance when translating long input sequences\cite{bahdanau2014neural}.

There are several reasons for this, but the main one is that the encoder is expected to be able to encode all the information of the source sentence in a single vector; in practice, especially for long input sentences, training such models with algorithms that back-propagate gradient updates can be challenging. For this reason, the \textit{attention} mechanism was developed. The main idea behind it is to extend the decoder with an attention layer that can access the entire sequence of encoder hidden states $h = (h_0, ..., h_k)$. The attention layer can then, at every decoding timestep $i$, compute an attention vector $\alpha_i$, whose values are normalized to add up to 1, and which constitute a series of weights that are used to compute the decoder's \textit{context vector} $c_i$, defined as:

$$c_i = \sum_{j=0}^{k} \alpha_{ij}h_j. $$


Intuitively, at every decoding timestep $i$, the corresponding context vector $c_i$ will change to reflect the greater (or lesser) relevance of some components of $h$ for computing the decoder output at that timestep. While the final encoder state $h_k$ is still fed into the decoder as an initial hidden state, the decoder is no longer wholly dependent on it, and has access to a much richer context thanks to the attention layer, which is trained with the rest of the model. Moreover, the existence of the attention layer can give additional insight into the inner working of the model. At evaluation time, the attention vectors can be collected and used to visualize what parts of an input have been focused on by the model when generating a certain output.

\section{Methods}\label{sec:arch}

\subsection{Problem Formulation}

The data used for this work comes from $4$ different signals from the TCV tokamak: the photodiode (PD), plasma current (IP), diamagnetic loop (DML), and interferometer (FIR). A more thorough description of those signals can be found in \cite{matos2020classification}. As in that work, we are generically interested in finding, for a given temporal sequence of measurements $x_t$, with $0 < t \leq N$ (which constitute a single shot), the most likely sequence of plasma confinement mode  $\hat{z}_{1:N}$  that explain the observations $x_{0:N}$.

The approach proposed in this paper does this in two parts. On the one hand, a model is trained to estimate the joint probability distribution of the sequence of plasma modes $p(z_1, z_2, ..., z_N | x_{0:N})$ for a given shot. Second, an algorithm finds a sequence $\hat{z}_{1:N}$, drawn from the joint distribution, with high probability. Formally, the task is to find:

$$\hat{z}_{1:N} = \argmax_{z_{1:N}} p(z_1, z_2, ..., z_N | x_{0:N}). $$

with $z_t \in Z$ and $Z: \{'Low', 'Dither', 'High'\}$, and $z_0 = \{'Low'\}$, since any shot is assumed to begin in L-mode.

In practice, in a real-time environment, we do not possess the entire sequence of measurements (the whole shot), but rather, only the signal values up to a certain point in time $t$. Thus, one of our requirements is to find a sequence of high probability up until $t$ while looking only at past measurements. For this task, a simple recurrent neural network (RNN) model can be used\cite{matos2020classification}. However, such RNN models, when making a decision, rely only on the input data and their own internal state. They cannot take their own past outputs into account when making a decision, and therefore, are limited to producing a single point-like estimate for the output sequence of confinement states. In contrast, sequence-to-sequence models can explicitly take their own past outputs into account when making a decision. This way, at time $t$, a sequence-to-sequence model does not decide on a single output for $t$, but rather, it decides on the entire sequence of plasma mode evolutions up to that point in time; that is, the model computes the distribution $p(z|x)$ as:

$$p(z_1, z_2, ..., z_N | x_{0:N})=p(z_1|x_{0:1}, z_0) p(z_2| x_{0:2}, z_{0:1})  ... p(z_N | x_{0:N}, z_{0:N-1}) = \prod_t p(z_t|x_{0:t}, z_{0:t-1}),$$

where $p(z_t|x_{0:t}, z_{0:t-1})$ denotes the probability of observing mode $z$ at time $t$, given the sequence of observed signals $x$ from time $0$ to time $t$, and the sequence of outputs until $t$. It is the additional conditioning on past outputs that allows the sequence-to-sequence model to approximate the full joint distribution $p(z|x)$. 

One caveat of the sequence-to-sequence model architecture is that it requires that the model observe windows, or subsequences, of the input data (up until time $t$) of fixed size. This means that in a real-time environment, for most values of $t$, a sequence-to-sequence model has a delay when computing $p(z_t|x_{0:t})$. This delay corresponds to a pre-defined size of the signal windows that the model receives, which therefore must be minimal. 

With a sequence-to-sequence model, using the notation above, finding a sequence of high probability means finding:

\begin{equation}
    \hat{z}_{1:N}= \argmax_{z_{1:N}} \prod_t p(z_t|x_{0:t}, z_{0:t-1}).
\end{equation}
\label{eq:sequence_max}

The task of finding samples of high probability from the distribution is done, in the case of this work, with a beam search algorithm. Because in our setting the sequences have potentially thousands of timesteps, computing their individual likelihoods using products as in Equation \ref{eq:sequence_max} would yield numerically unstable results; thus, the beam search uses the logarithm of the probabilities:

\begin{equation}
\eqalign{\hat{z}_{1:N}= \argmax_{z_{1:N}} \log(\prod_t p(z_t|x_{0:t}, z_{0:t-1})) \cr
= \argmax_{z_{1:N}} \sum_t \log p(z_t|x_{0:t}, z_{0:t-1}),} 
\end{equation}
\label{eq:log_sequence_max}

which allows it to use sums instead, and to look for the sequence whose log-probability is greatest.

\subsection{Data engineering}\label{sec:dataengineering}

Events such as, for example, the LH transition can be roughly pinpointed in a signal time trace. However, it is difficult to specify precisely, on a consistent basis, at which exact point in time the transition happens; for example, in some shots, the transition might be quite sudden, whereas in others, the transition signatures in the data can be more spread out over time. Indeed, the typical time that a TCV shot takes to make an HL transition is on the order of 1 ms. Considering also that the sampling frequency of our signals is 10 kHz, this translates to an intrinsic uncertainty for the label determination of at least 10 time steps. So, for example, one expert might determine, for a shot, the start of the H mode at the point in time where the PD signal starts dropping, whereas another expert might claim that the H mode actually only starts at the point where the signal has already stabilized (perhaps 1 ms later). While the difference may sound trivial, this can become problematic in a supervised learning task such as the one we face in this work. In principle, if the amount of training data is sufficiently large, small inconsistencies in labels (such as variations of 1 ms in the localization of transitions) will tend to be averaged out by a classifier. Nevertheless, these mismatches can produce instabilities during training and can ultimately degrade performance. 

\begin{figure*}[h!]
    \centering
        \includegraphics[scale=0.75, trim=180 270 180 210, clip]{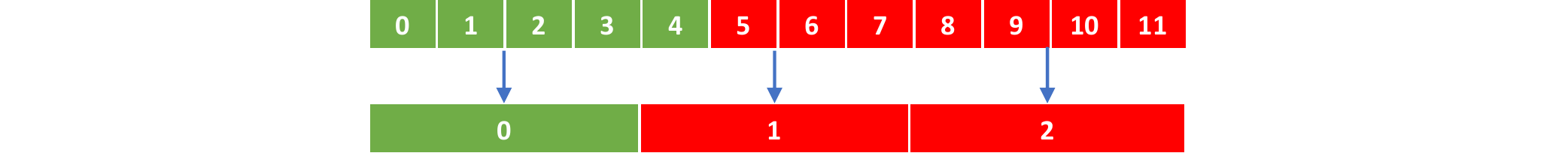}
        \caption{Representation of the computation of a block of target labels, where each bottom block corresponds to $4$ source timesteps. The green color indicates a label of L (Low confinement) while red represents H mode. A majority of the labels in the source timesteps 4 - 7 are in H mode, so the corresponding block at target timestep 1 is labeled as H. }
        \label{fig:block}
\end{figure*}

For that reason, one of the steps we took in this work is to reduce the temporal resolution of the sequence-to-sequence model's outputs. This can be done thanks to the model's architecture, which allows for a mismatch between the size of the input and output sequences. We do this by grouping the existing labels (i.e., sequences of shot classifications) in our dataset into \textit{blocks} of a fixed size. In the pre-processing stage, each of those blocks is mapped to a certain plasma confinement mode (L, D or H); this mapping is done by computing the source label with the highest number of occurrences within that block (see Figure \ref{fig:block}). Our expectation is that this decrease in temporal resolution will yield better performance in both training and inference time, at a minimal cost to the physical validity of the results.

\subsection{Model architecture}

Our model's architecture is based on existing architectures for neural machine translation, in particular, the one proposed in \cite{luong2015effective}. The architecture consists of an encoder-decoder model with attention, which we detail in the following paragraphs.

Unlike most work with language translation, where one receives discrete units (words) as inputs, in our case, the inputs to the model are the continuous signal time-traces from TCV shots. In those time-traces, one expects to find not only localized, spatial correlations in the data --- for example, a sudden drop in the PD signal --- but also long-term contextual correlations, namely, which modes a shot may have been in in the past. For that reason, the encoder in our model consists of a convolutional recurrent neural network, much like the one used in\cite{matos2020classification}. We made slight adjustments, namely in the number of convolutions used, but otherwise preserved the architecture, and used long short-term memory (LSTM) units for the recurrent layers. The inputs fed to the encoder are sequences of overlapping windows, which slide across the signal time-series. These windows were defined to have a size of $40$ source timesteps, with a stride between windows of $10$ timesteps. This last detail, in particular, means that the full sequence-to-sequence model has approximately 10 times fewer network parameters than the convolutional LSTM in previous work. 

The convolutional layers are trained to find local correlations in the windows, while the recurrent layers, based on the output of the convolutions, keep track of long-term correlations in the input sequences. For example, the convolutional layers can detect a local shape in a signal window indicating a possible L-H transition, and the recurrent layers then use that, as well as their stored information about the current plasma mode (i.e. the long-term dependency), to determine whether a transition indeed has occurred. 

In \cite{matos2020classification}, this Convolutional LSTM was used to directly map the input sequence of measurements, $x_t$, into a sequence of outputs, $z_t$, indicating a plasma confinement mode at a particular point in time (see Equation \ref{eq:sequence_max}). In this work, the task of this particular submodel is instead to produce an encoded summary of $x_t$, which is stored in its internal hidden state vector $h$. During both training and inference time, we feed the encoder not with entire shots, but rather, subsequences of signals drawn from the shots. There are several reasons for this. On one hand, in recurrent neural networks, gradients can vanish when being backpropagated through time, which can be particularly problematic when working with long sequences; training with smaller subsequences mitigates this problem. On the other hand, the existing data is imbalanced with regard to the labels; for example, Dithers tend to be much less frequent than L and H modes. Using subsequences allows us to feed the entire model, during the training process, with a more balanced number of samples for each class, thus preventing a potential source of biased results. Finally, any future usage of the methods proposed in this paper for real-time plasma data analysis implies, by definition, that only information until a particular point in time, and not the entire shot, is available for classification. 

In training time, the subsequences are drawn uniformly (with respect to the three classes) from our existing data ensemble. In inference time, for any given shot, the subsequences are drawn and fed to the encoder consecutively. In both cases, the encoder's state is reset each time a new subsequence is fed to it, which means that the context vectors only hold information about the current subsequence under consideration. 

\begin{figure*}[h!]
    \centering
        \includegraphics[scale=0.75, trim=300 280 300 60, clip]{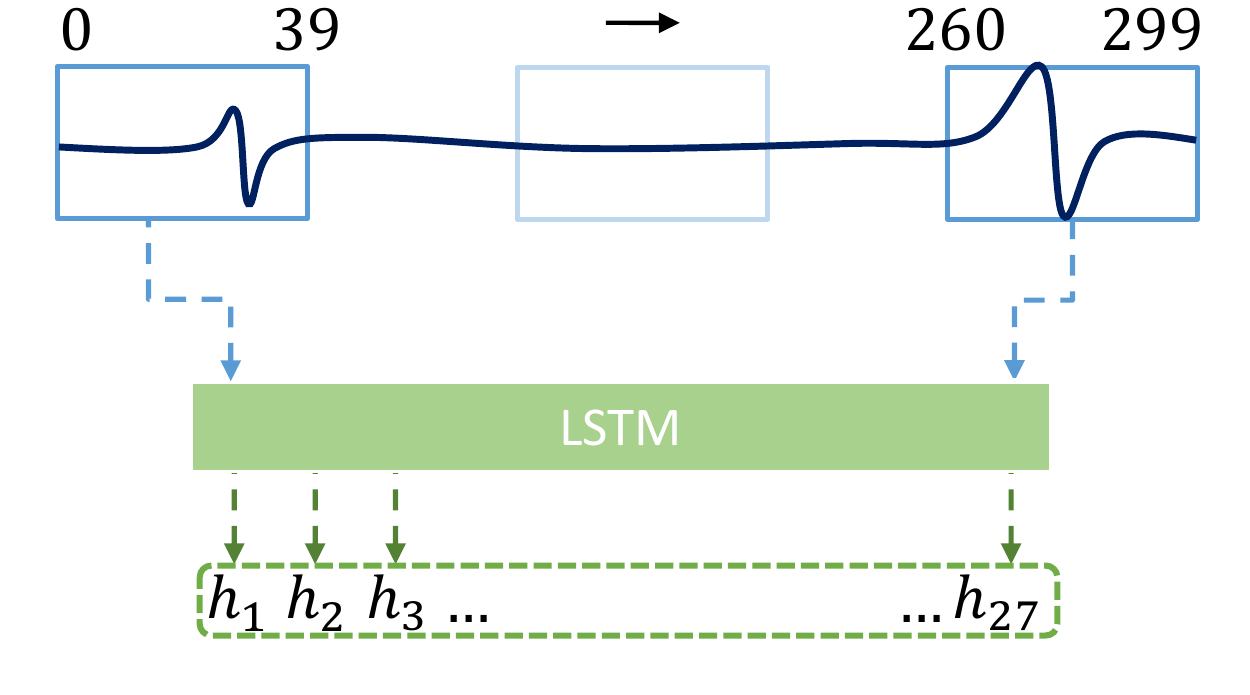}
        \caption[Illustration of the encoder architecture]{Illustration of the encoder architecture and how a single subsequence is processed and encoded in the encoder hidden state vector $h$. The dark blue line represents a subsequence of a signal timeseries. The sliding windows are in light blue.}
        \label{fig:conv_encoding}
\end{figure*}

In practice, we defined the subsequences drawn for training and inference to have a size of 300 source timesteps (this contrasts with a typical shot size of, at least, $10000$ timesteps at a $10kHz$ sampling rate). This value also corresponds to the delay that the model would have in a real-time setting. With a window size of 40 and a window stride of 10, these 300 timesteps are fed to the convolutional layers as sequences of 27 convolutional windows (further shrinking the total size of the sequence fed to the RNN), with window $1$ observing $x_t$ from $t=0$ to $t=39$, and window 27 observing $x_t$ from $t=260$ to $t=299$. The convolutional LSTM processes these windows to produce the hidden state vector $h$, which has a length of $27$ elements and is fed to the decoder. An illustration of this can be found in Figure \ref{fig:conv_encoding}.

The task of the decoder is to approximate a probability distribution $p(z|x)$ of plasma confinement modes, subject to the summary given to it by the encoder. Our decoder is composed of a recurrent neural network (specifically, an LSTM layer), an attention layer, and a series of dense layers. 

Each individual element of the output sequence is processed in a single timestep, with each target timestep corresponding to a single block of classifications. We defined the blocks as having a size of $10$ - that is, each target timestep yields a single classification for $10$ source timesteps.

\begin{figure*}[h!]
    \centering
        \includegraphics[scale=0.65, trim=250 110 250 100, clip]{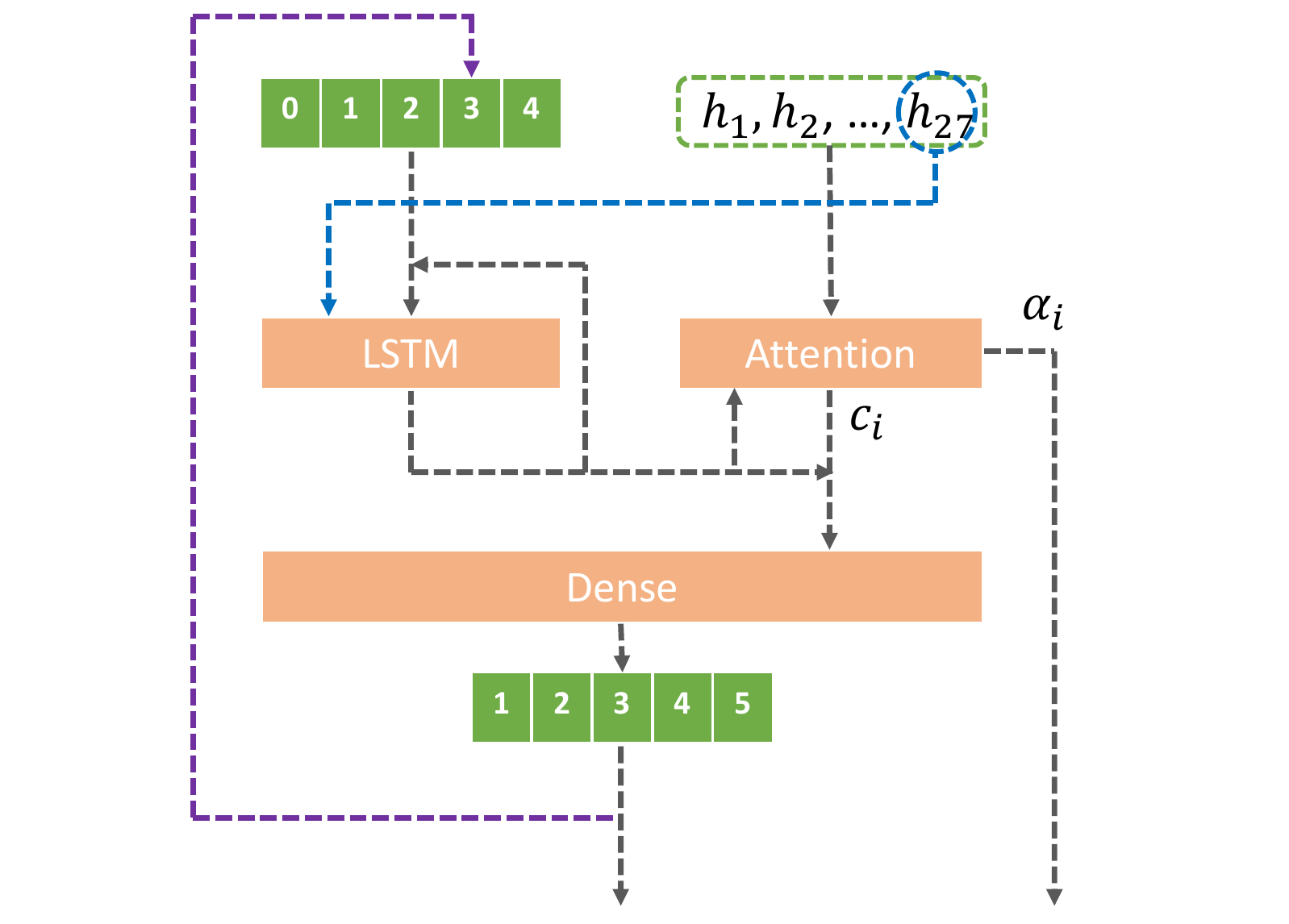}
        \caption{Schematic representation of the decoder's architecture. Represented here is the sequence of operations carried out by the decoder to produce the output distribution of plasma confinement modes at a decoding timestep \textit{t}. Gray arrows denote data flows at $t$. The purple arrow denotes the autoregressive feeding of the output at $t+1$. The blue arrow denotes the initial setting of the decoder state to the last encoder state. Joining arrows denote concatenation, $\alpha_i$ and $c_i$ are the attention weights and the context vector, respectively.   }
        \label{fig:decoder}
\end{figure*}

In the first decoding timestep of each new subsequence, the decoder's initial hidden state is set to the decoder's final hidden state, $h_k$, for that subsequence. At each decoding timestep, the decoder receives as input its own last processed output (plasma confinement mode $z_{t-1}$), and the last output from its own LSTM cell; these are concatenated as suggested in \cite{luong2015effective} and fed to the decoder's LSTM, which also updates its own internal state. The output of the LSTM is then concatenated to the output of the attention layer (which receives the entire encoder hidden state vector $h$), and fed through a series of dense layers to produce the final decoder output, which is a vector whose entries add up to 1 and which, individually, represent the computed probability of a given plasma confinement mode $z_t$ (see Figure \ref{fig:decoder}). The decoder's LSTM is built with a latent dimensionality of $32$ units, that is, we process each target timestep with $32$ LSTM cells. In terms of the attention layer, in \cite{luong2015effective}, several different mechanisms for computing the alignment scores are proposed; we opted for the general form described in that paper. 

\begin{figure*}[h!]
    \centering
        \includegraphics[scale=0.6, trim=250 150 250 110, clip]{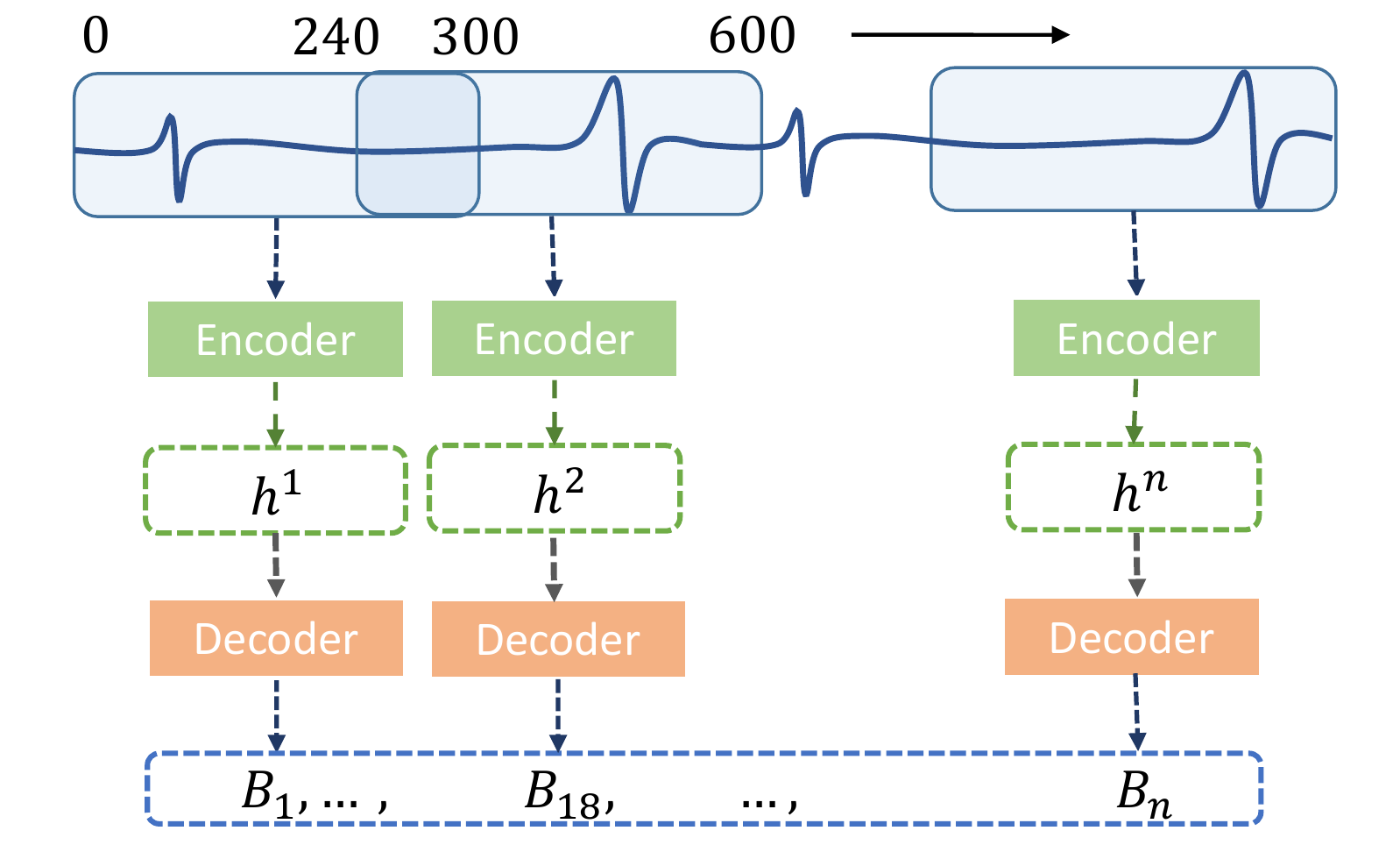}
            \caption[Representation of the full sequence-to-sequence model's architecture.]{Representation of the full sequence-to-sequence model's architecture. Notice how the subsequences (translucid blue) overlap with each other. The encoder's state vectors, $h^k$ (where $k$ is the current subsequence), are fed to the decoder, which produces blocks of outputs.  }
        \label{fig:model_arc}
\end{figure*}

One of our considerations when designing the decoder was to take into account how a human expert would label the signals. Our intuition was that, when looking at a time point of a shot, a labeler always takes into consideration information around that point --- namely, the events happening immediately before and after. For that reason, we designed the decoder such that, for each incoming context vector (which encodes information regarding $300$ source time steps), the decoder produces a sequence of 18 blocks of labels (with a block size of $10$, this corresponds to $180$ source time steps), which are the classifications for source subsequence steps $[60:240]$. The idea is that the extra information present in the remaining source time steps would help improve the classification results. In evaluation time, this setup requires that the consecutive subsequences drawn from a shot overlap with each other, so that an output sequence can be produced for the entire shot. This also leads to the loss of the initial and final $6$ target classification blocks of a shot (see Figure \ref{fig:model_arc}), but we consider this to have no bearing on our results.

In evaluation time, the decoder always produces a distribution of possible plasma confinement states whose probabilities add up to 1. One possibility would be to use a greedy aproach and simply take, at each target timestep, the plasma state for which the output probability is highest, and feed that state at the next decoding timestep. This would yield a possible solution (i.e., a single sequence of plasma confinement states), but there would be no guarantee of it being optimal. For that reason, we use a beam search algorithm to traverse the tree structure of possible solutions (different sequences of $z$), which allows for obtaining samples closer to the optimal $\hat{z}$. This is done by, at each target timestep, expanding the search tree for all previous outputs, and not just for the output of highest probability. Then, for each previous output $z_{t-1}$ under consideration, the conditional probability, and the log-likelihood, defined in equation \ref{eq:log_sequence_max} are computed for the current timestep $t$. Once all target timesteps have been processed, the value of $z$ for which the likelihood is highest is returned. 

\begin{figure*}[h!]
    \centering
        \includegraphics[scale=0.5, trim=100 50 130 10, clip]{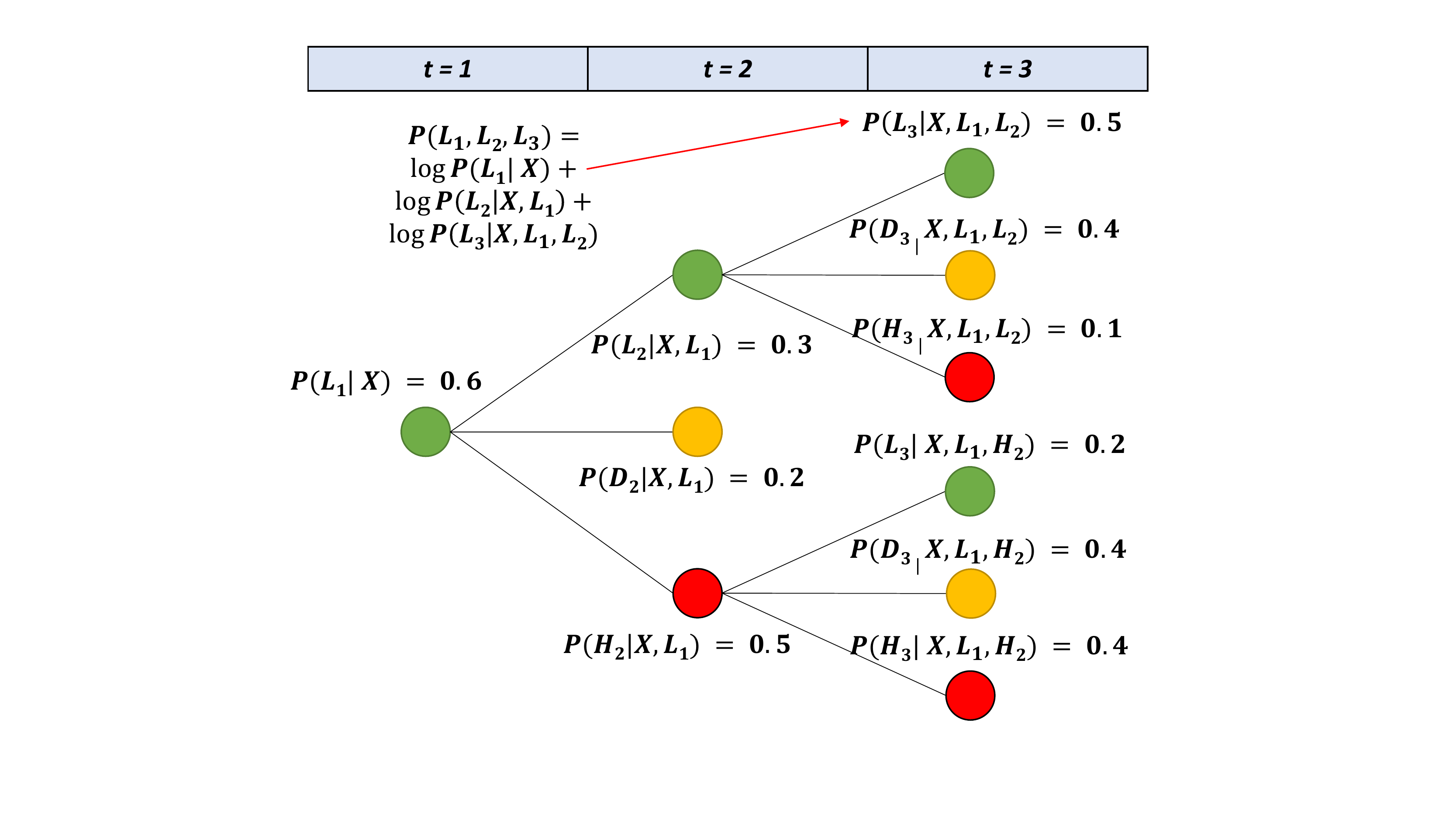}
        \caption{Illustration of the beam search algoritm in the first 3 target timesteps. X denotes the context vector and the final encoder state which condition the output. In timestep 1, the decoder computes the probability of a path starting in L mode, and we manually set that to be the only path to be expanded. In timestep 2, the conditional probabilities of three paths are computed; the paths that have L or H are expanded.}
        \label{fig:beam_search}
\end{figure*}

Naturally, expanding the full tree at every timestep would quickly become unwieldy, owing to the large number of different paths to be processed. On the other hand, expanding a single beam at each timestep would be equivalent to a greedy search. For these reasons, we defined the beam search to have a maximum width of $20$, i.e., at every step, only the $20$ paths with the highest log-probability are expanded, while the rest are discarded. In addition, we encoded in the beam search a rule for expanding only those paths that start in L-mode, which simplifies the search; an illustration of this can be seen in Figure \ref{fig:beam_search}.

\subsection{Dataset Preparation}

For the work in \cite{matos2020classification}, a total of $54$ shots were used for training and validating the proposed models. In this work, we carried out a more careful treatment of the dataset preparation with respect to the previous publication. In the first place, the selection of the discharges for training and testing was done in order to cover as  exhaustively as possible the space of the plasma confinement modes in TCV, accounting for the different temporal evolutions of the plasma. Using a Dynamic Time Warping (DTW) algorithm, we measured the similarity between pairs of temporal sequences and assigned them to a given group, based on a similarity measure. The desired number of groups was obtained by applying a Hierarchical Clustering algorithm to univariate time sequences corresponding to the entire plasma discharges. A total of 293 discharges were selected and processed through the DTW, setting the number of clusters to 100. From each of the 100 clusters, shots were extracted and classified as an interesting (or not) shot from the physics point of view, as far as our problem (i.e., the presence of L/D/H transitions) is concerned. Some clusters were discarded since they consisted of disruptions without even achieving an H mode confinement state, while others presented technical issues in the ramp-up phase before reaching the stationary phase. Limiting ourselves to the interesting shots and maximizing the number of clusters where a given shot arose from, 88 discharges were selected for further validation (ground truth determination). For the latter, instead of having different validations by different labelers for a particular shot, a consensus on a common convention between two experts was established to determine the label of each time step for all shots. A detailed revision of the different transitions was performed with particular attention to the presence of short transitions on the order of 2ms. The outcome was a unique, consistent, ground truth per shot.
A test set to evaluate the final results of the model was carefully determined and fixed during all the experiments. A total of 27 shots were selected, each extracted from a different cluster. Out of 27 shots, 17 shots were “unpolluted” cases (without the presence of type III ELMs), while the others 10 were special “noisy” discharges with type III ELMs. The proportion between the noisy and “unpolluted” discharges in the test set followed approximately the same proportion as in the complete dataset.

\textbf{\section{Results}\label{sec:results}}

We begin this section with a direct comparison between this sequence-to-sequence model, and the convolutional LSTM used in\cite{matos2020classification}. To that end, we trained and tested the old model, preserving all the original architecture and hyperparameters, with the new train and validation data (shots and labels) compiled for this work. Table \ref{tab:kappa_seq2seq_old_data} shows the results. As in the previous work, we performed the evaluation using Cohen's Kappa-statistic coefficient\cite{landis1977measurement}, which gives an indication of the match between two sets of categorical data (with a score of $1$ for a perfect match and $0$ for no match). In our case, it reflects the match between the models' outputs, and the labeled data. We computed the score on a per-class basis and also on a weighted mean basis, in order to indicate whether the classifications produced by the sequence-to-sequence model match the data's labels. We designed the model with Tensorflow\cite{tensorflow2015-whitepaper}, and ran it on an NVIDIA Quadro RTX 5000 GPU.

\begin{table}[h!]
    \begin{center}
        \begin{tabular}{cccccc}
         &  & L & D & H & Mean \\
         \hline
         \multirow{2}{*}{$\kappa$ scores} & Train & 0.98 & 0.91 & 0.98 & 0.98 \\
         & Test  & 0.92 & 0.78  & 0.91 & 0.9 \\
        \end{tabular}
    \end{center}
    \vspace{-5mm}
    \caption[$\kappa$-statistic scores for the Conv-LSTM model on the data used for this work.]{$\kappa$-statistic scores for each plasma mode and as a mean, on training and test data, for the Conv-LSTM model from\cite{matos2020classification} on the data used for this work. }
    \label{tab:kappa_seq2seq_old_data}
\end{table}

A comparison between the results in Table  \ref{tab:kappa_seq2seq_old_data} and those described in \cite{matos2020classification} shows that the current dataset already improved the capacity of the old model. Nevertheless, it was still underperforming, particularly on dithers; even on training data, the mean dither score was 0.9.

\begin{figure}[h!]
    \centering
    \begin{subfigure}{0.45\textwidth}
        \centering
        
        \includegraphics[width=\textwidth,
        scale = 0.55, 
        trim= 0 0 0 0,
        clip]{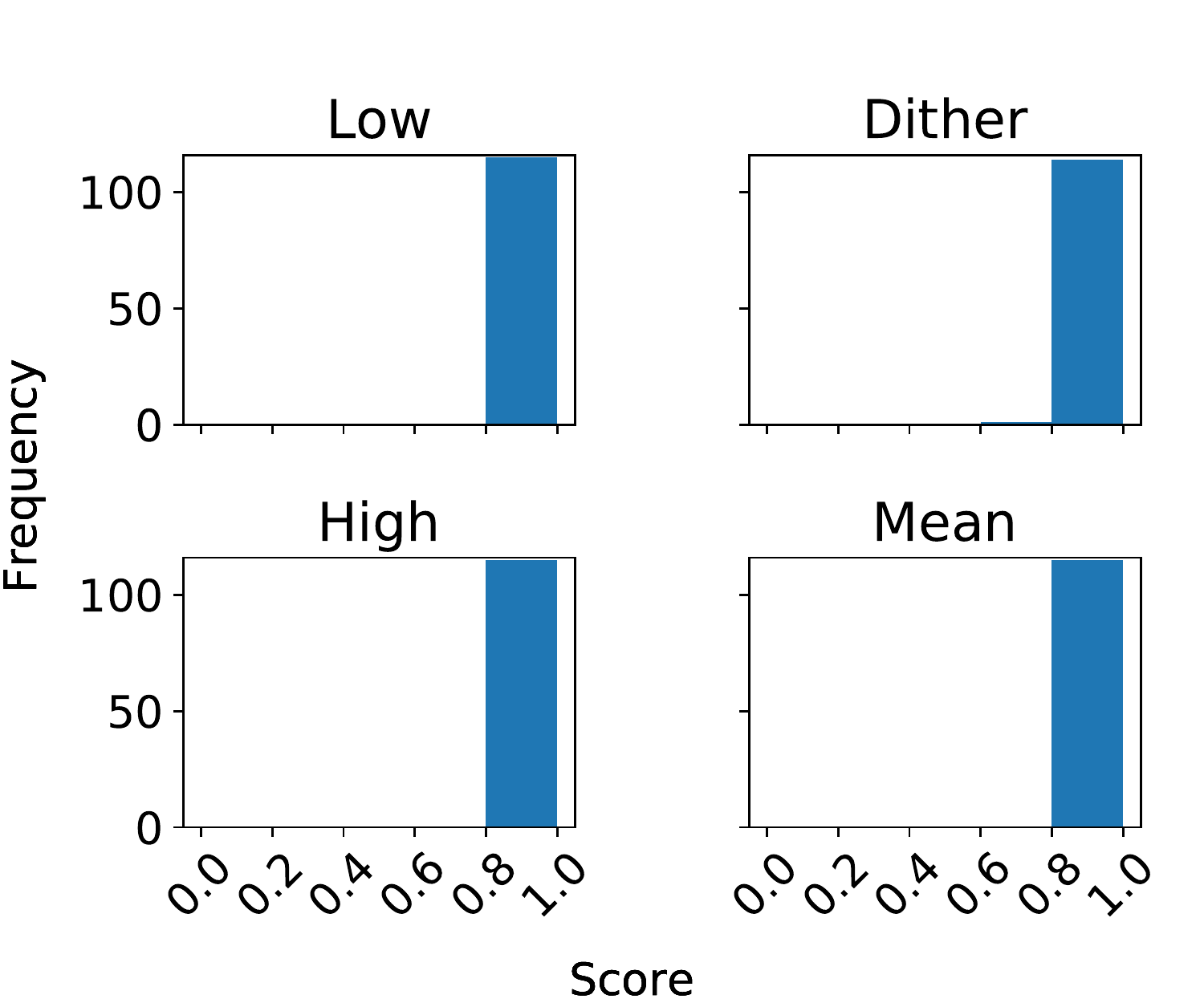}
        \caption{Training data.}
        \label{subfig:seqtoseq_train_histo}
    \end{subfigure}%
    ~
    \begin{subfigure}{0.45\textwidth}
        \centering
        
        \includegraphics[width=\textwidth,
        scale = 0.55, 
        trim= 0 0 0 0,
        clip]{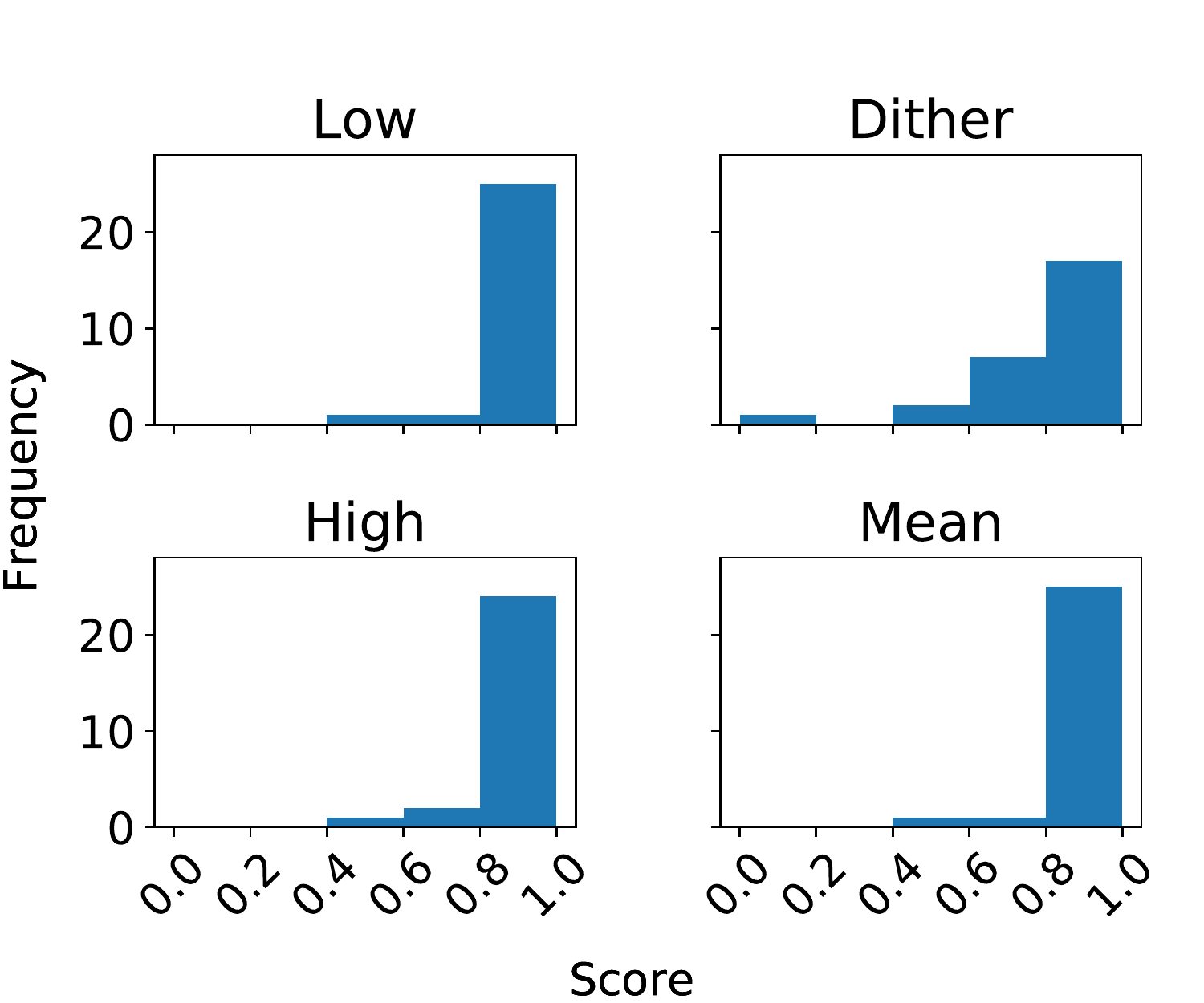}
        \caption{Test data.}
        \label{subfig:seqtoseq_test_histo}
    \end{subfigure}%
    \vspace{-2mm}
    \caption{Distribution of the $\kappa$-statistic score on a per-shot basis}    \label{fig:cnn_histograms}
\end{figure}

We then ran the new sequence-to-sequence model. The results on both the train and test sets can be seen in Table \ref{tab:kappa_seqtoseq}. They were obtained by training the network for 150 epochs, with each epoch consisting of 128 batches of data, and each batch consisting of 128 data samples, i.e., uniformly sampled subsequences of each of the existing classes drawn from the training shots. We downsampled the source labels to the same temporal resolution as the model's output blocks to compute the scores. Figures \ref{subfig:seqtoseq_train_histo} and \ref{subfig:seqtoseq_test_histo} show the distribution of the scores on a per-class basis, for train and test data, as well as a weighted mean value, taking into account the relative frequencies of each class in the labels. 

\begin{table}[h!]
    \begin{center}
        \begin{tabular}{cccccc}
         &  & L & D & H & Mean \\
         \hline
         \multirow{2}{*}{$\kappa$ scores} & Train & 0.99 & 0.99 & 0.99 & 0.99 \\
         & Test & 0.94 & 0.86 & 0.96 & 0.94 \\
        \end{tabular}
    \end{center}
    \vspace{-5mm}
    \caption{$\kappa$-statistic scores for each plasma mode and as a mean, on training and test data, for the sequence-to-sequence model. }
    \label{tab:kappa_seqtoseq}
\end{table}

\section{Discussion}\label{sec:discussion}

In this section, we discuss the results in further detail, in particular, the cases where the sequence-to-sequence model's performance was poorer.

With regards to the training data, the classification was excellent in all cases; all shots achieved scores, both on a per-mode basis and as a mean, above 0.8. The lowest mean train score for a shot was $0.965$, while the lowest scores for L, D and H modes were, respectively, $0.96$, $0.8$ and $0.98$. Nevertheless, for all modes, the mean score was $0.99$, a result that indicates that the sequence-to-sequence model has the capacity to learn the underlying correlations in the data to make accurate predictions. 

With regard to the validation data, the scores are slightly lower. Table \ref{tab:seq2seq_results} shows the detailed breakdown of the scores for all shots where at least one mode had a score lower than $0.8$. Some of those shots have low scores on more than one class: for example, $\#42197$ had a score lower than 0.8 for both L and D modes. Notice how, even though there are more shots with lower dither scores, the overall mean values are in most cases above $0.8$; this is due to the fact that dithers are rather less frequent in the data than L and H modes.

\begin{table}[h]
    \begin{center}
    \item[]
        \begin{tabular}{|c|c|c|c|c|c|c|c|}
            \hline
            \multirow{2}{*}{Shot ID} & \multicolumn{2}{c|}{L} & \multicolumn{2}{c|}{D} & \multicolumn{2}{c|}{H} &  \\ \cline{2-8} 
             & Fraction & Score & Fraction & Score & Fraction & Score & Mean \\ \hline
            42197 & 0.57 & 0.5 & 0.35 & 0.66 & 0.08 & 0.46 & 0.55 \\ \hline
            61057 & 0.5 & 0.72 & 0.04 & 0.68 & 0.46 & 0.73 & 0.72 \\ \hline
            61274 & 0.69 & 0.84 & 0.12 & 0.7 & 0.19 & 0.96 & 0.84 \\ \hline
            32911 & 0.52 & 0.84 & 0.28 & 0.79 & 0.20 & 0.97 & 0.85 \\ \hline
            61043 & 0.78 & 0.92 & 0.13 & 0.69 & 0.1 & 0.76 & 0.87 \\ \hline
            45105 & 0.42 & 0.91 & 0.02 & 0.52 & 0.56 & 0.94 & 0.92 \\ \hline
            34309 & 0.14 & 1 & .03 & 0.8 & 0.83 & 0.96 & 0.96 \\ \hline
            33459 & 0.8 & 0.98 & 0.01 & 0.5 & 0.19 & 1 & 0.98 \\ \hline
            64376 & 0.92 & 0.9 & .01 & 0.73 & 0.08 & 1 & 0.98 \\ \hline
            30268 & 0.24 & 1 & 0 & 0 & 0.76 & 1 & 1 \\ \hline
        \end{tabular}
    \end{center}
    \vspace{-5mm}
    \caption{Kappa statistic scores for shots with at least one mode whose score was lower than $0.8$. The ``fraction'' column indicates what percentage of the labels were labeled in a particular state for that shot.}
    \label{tab:seq2seq_results}
\end{table}

\begin{figure*}[h!]
    \centering
        \includegraphics[scale=0.42, trim=55 40 35 40, clip]{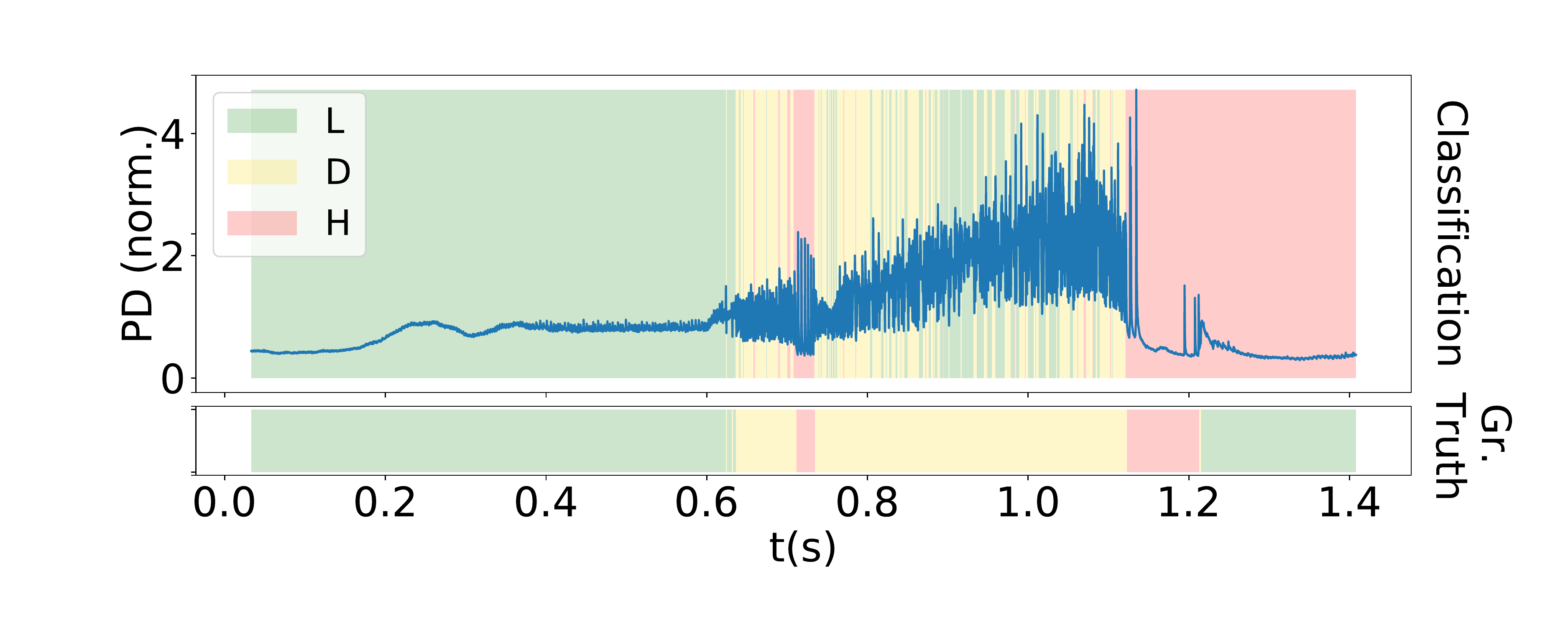}
        \caption[Detection results for shot $\#42197$.]{Detection results for shot $\#42197$. The blue line denotes the photodiode signal, while the solid background color denotes the confinement states. }
        \label{fig:results_seq2seq_42197}
\end{figure*}

Figure \ref{fig:results_seq2seq_42197} shows the classification results for shot $\#42197$, together with the ground truth (label), shown in the lower panel of the Figure (we show only the photodiode signal values for ease of comprehension). The classification has two major errors: the non-detection of a switch to L mode at the end of the shot, around $t=1.2s$, and the rapid oscillation between L and D modes from approximately $t=0.8s$ to $t=1.1s$. This oscillation in particular is questionable because this dithering phase presented particularly odd fluctuations that were not a common behavior in our dataset. 

\begin{figure*}[h!]
    \centering
        \includegraphics[scale=0.42, trim=55 40 35 40,clip]{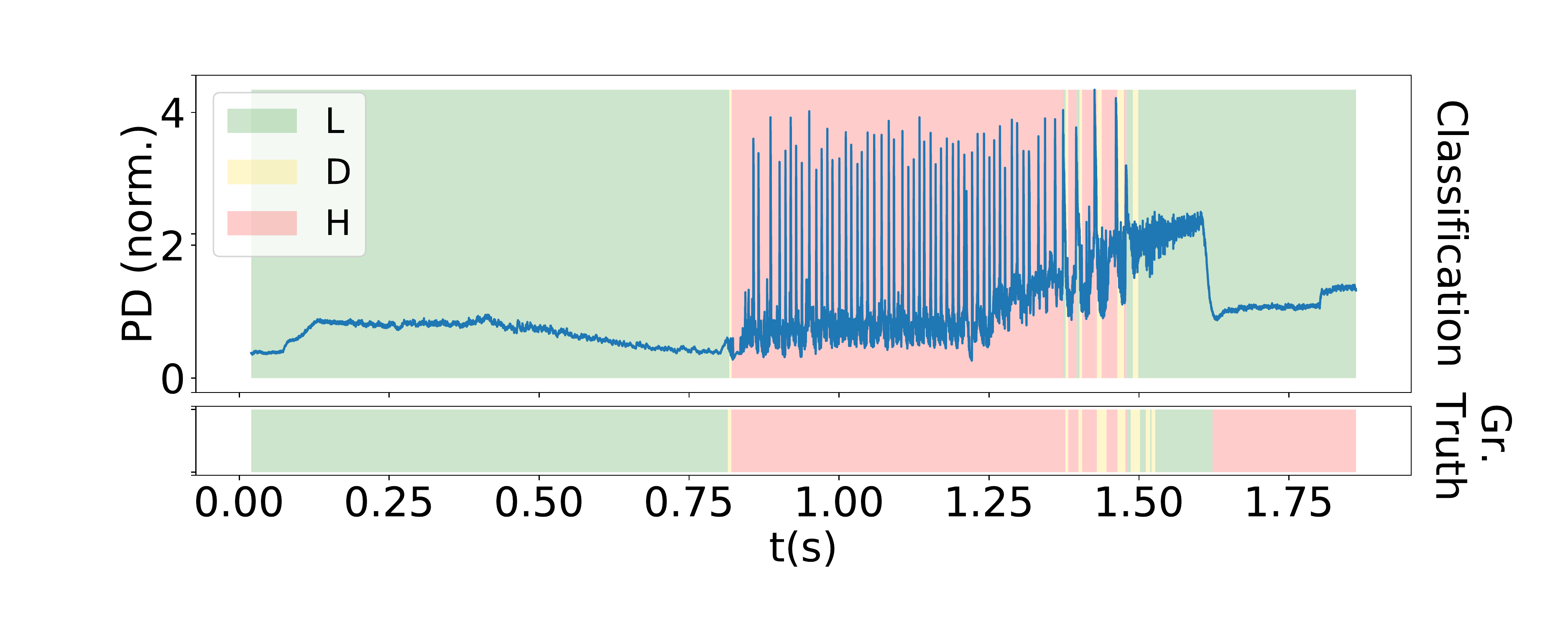}
        \caption[Detection results for shot $\#61057$.]{Detection results for shot $\#61057$.}
        \label{fig:results_seq2seq_61057}
\end{figure*}

For shot $\#61057$, two short dither bursts near $t=1.5s$ are missed, as well as the final transition back into H mode near $t=1.6s$ (see Figure \ref{fig:results_seq2seq_61057}). 

\begin{figure*}[h!]
    \centering
        \includegraphics[scale=0.42, trim=55 40 35 40,clip]{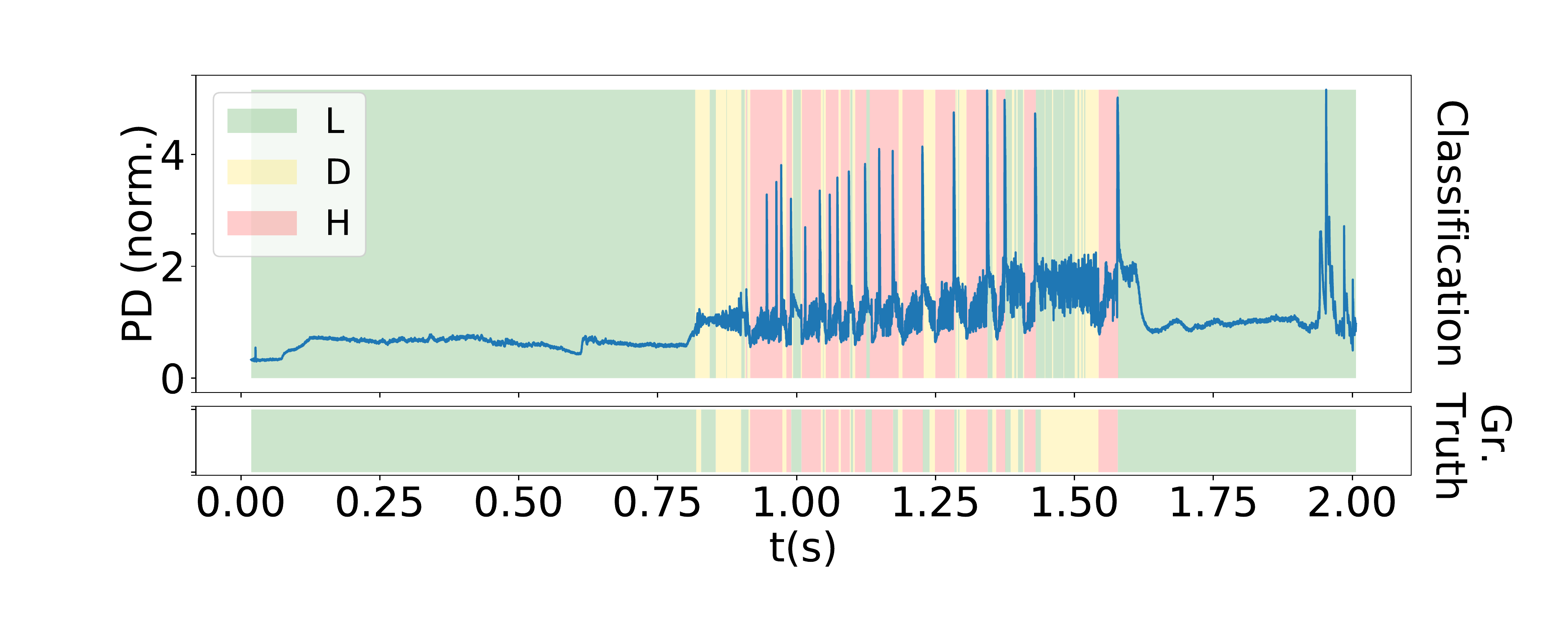}
        \caption[Detection results for shot $\#61274$.]{Detection results for shot $\#61274$.}
        \label{fig:results_seq2seq_61274}
\end{figure*}

For shot $\#61274$, the overall classification score is already above $0.8$; the model's largest mistake is in incorrectly classifying a region around $t=1.5s$ as L mode, which explains both the lower score for that mode and for dither (see Figure \ref{fig:results_seq2seq_61274}).

\begin{figure*}[h!]
    \centering
        \includegraphics[scale=0.42, trim=55 40 35 40,clip]{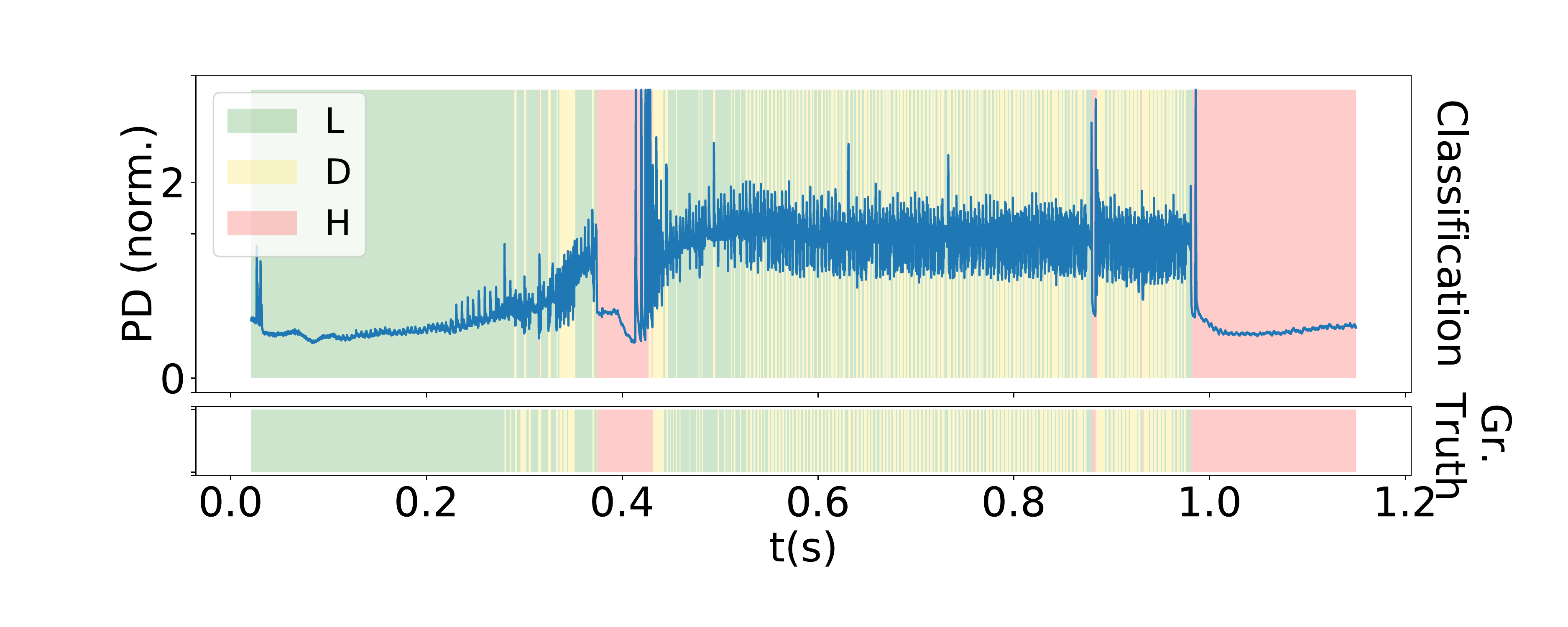}
        \caption[Detection results for shot $\#32911$.]{Detection results for shot $\#32911$.}
        \label{fig:results_seq2seq_32911}
\end{figure*}

In shot $\#32911$, the lower scores for dither and L mode are also due to rapid fluctuations between the two modes; we think that a big factor behind the lower score is the accumulation of many small mismatches between the labels and the classifications (see Figure \ref{fig:results_seq2seq_32911}).

\begin{figure*}[h!]
    \centering
        \includegraphics[scale=0.42, trim=55 40 35 40,clip]{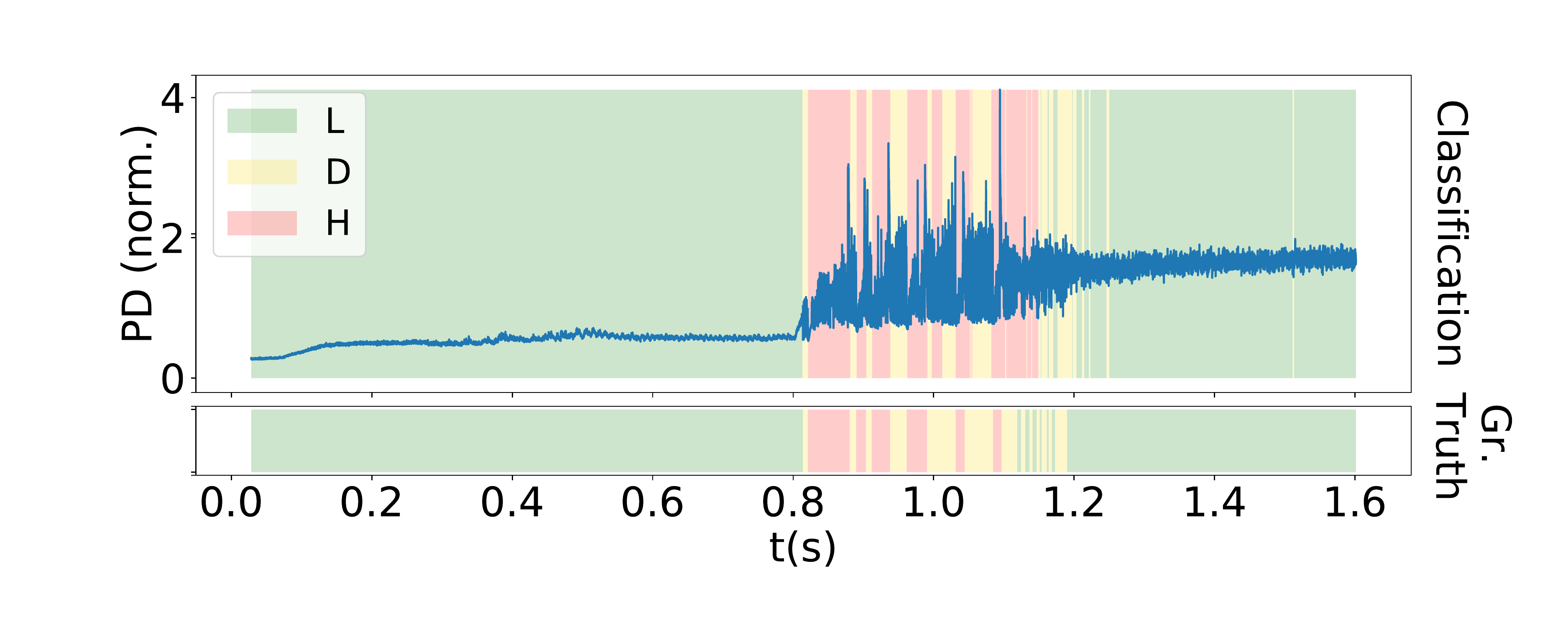}
        \caption[Detection results for shot $\#61043$.]{Detection results for shot $\#61043$.}
        \label{fig:results_seq2seq_61043}
\end{figure*}

Shot $\#61043$ contains several switches between H mode and dithering; the model incorrectly classifies some of these  (see Figure \ref{fig:results_seq2seq_61043}). 

We consider the remaining validation to be good classifications overall. We explain the occasionally lower dither scores by the fact that the labeled dither phases are very short (at most $3\%$ of any given shot); any mismatch, even if small, between the label and dither classification produces a low score.

\section{Conclusions}

This work developed a sequence-to-sequence neural network model with attention for automated classification of plasma confinement states at the TCV tokamak. Taking previous work \cite{matos2020classification} with convolutional recurrent neural networks for the same task as a baseline, our intuition was that one of the factors holding previous models back from achieving even better scores was the fact that vanilla RNNs cannot reason over different past outputs, and are limited to the signal observations. This contrasts with the way a human labeler would perform the same task; typically, a labeler will reason over different possibilities for past states before deciding on a label for a given timestep, which is a process that a sequence-to-sequence model can emulate. In addition, we also hypothesized that performance was limited by the lack of sufficient train data. Therefore, we extended the dataset used in previous work with more shots and chose a train/validation split that thoroughly represented the operational shot space. In addition, we took particular care with the labeling process, to ensure a high quality of the data.

Testing the convolutional recurrent model from \cite{matos2020classification} on the new dataset increased its scores on both validation and train data, suggesting that indeed the size and quality of the dataset was having an impact on the obtained results. Running the sequence-to-sequence model with the new data, we achieved excellent results on the train set (with a mean score of $0.99$ out of $1$), and very good results on the validation set (with a mean score of $0.94$ out of $1$). In this case, we conclude that the new sequence-to-sequence model that we propose indeed worked as we expected, given that it correctly learned all the training data. We explain the slightly lower validation score with the fact that several validation shots were particularly challenging to classify, even for a human labeler. Nevertheless, the results obtained indicate that the task of plasma confinement mode classification can be correctly carried out by deep learning algorithms, in particular the sequence-to-sequence model presented here. 

\section*{Acknowledgments}
\small
\noindent This work has been carried out within the framework of the EUROfusion Consortium and has received funding from the Euratom research and training programme 2014-2018 and 2019-2020 under grant agreement No 633053. The views and opinions expressed herein do not necessarily reflect those of the European Commission.

\clearpage

\section*{References}
\bibliography{bibliography}

\end{document}